\documentclass[
reprint,
superscriptaddress,
 amsmath,amssymb,
 aps,
prb,
]{revtex4-2}

\usepackage{graphicx}
\usepackage{dcolumn}
\usepackage{bm}
\usepackage{braket}

\usepackage[T1]{fontenc}    
\usepackage{lmodern}           
\usepackage{textcomp} 
\usepackage{xr}

\usepackage{color}
\usepackage{xcolor}

\begin{document}

\preprint{APS/123-QED}

\title{Mitigation of birefringence in cavity-based quantum networks using frequency-encoded photons}
\author{Chengxi Zhang}
\affiliation{Department of Physics, University of California, Berkeley, CA 94270, USA}

\author{Justin Phillips}%
\affiliation{Department of Physics, University of California, Berkeley, CA 94270, USA} 
\affiliation{Lawrence Berkeley National Laboratory, Berkeley, CA 94720, USA}

\author{Inder Monga}
\affiliation{Lawrence Berkeley National Laboratory, Berkeley, CA 94720, USA}

\author{Erhan Saglamyurek}
\affiliation{Department of Physics, University of California, Berkeley, CA 94270, USA}
\affiliation{Lawrence Berkeley National Laboratory, Berkeley, CA 94720, USA}

\author{Qiming Wu}
\email{qiming.wu@berkeley.edu}
\affiliation{Department of Physics, University of California, Berkeley, CA 94270, USA}
\affiliation{Lawrence Berkeley National Laboratory, Berkeley, CA 94720, USA}

\author{Hartmut Haeffner}
\affiliation{Department of Physics, University of California, Berkeley, CA 94270, USA}
\affiliation{Lawrence Berkeley National Laboratory, Berkeley, CA 94720, USA}

\date{\today}

\begin{abstract}
Atom-cavity systems offer unique advantages for building large-scale distributed quantum computers by providing strong atom-photon coupling while allowing for high-fidelity local operations of atomic qubits. 
 However, in prevalent schemes where the photonic state is encoded in polarization, cavity birefringence introduces an energy splitting of the cavity eigenmodes and alters the polarization states, thus limiting the fidelity of remote entanglement generation. To address this challenge, we propose a scheme that encodes the photonic qubit in the frequency degree-of-freedom. The scheme relies on resonant coupling of multiple transverse cavity modes to different atomic transitions that are well-separated in frequency. 
 We numerically investigate the temporal properties of the photonic wavepacket, two-photon interference visibility, and atom-atom entanglement fidelity under various cavity polarization-mode splittings and find that our scheme is less affected by cavity birefringence. Finally, we propose practical implementations in two trapped ion systems, using the fine structure splitting in the metastable D state of $\mathrm{^{40}Ca^{+}}$, and the hyperfine splitting in the ground state of $\mathrm{^{225}Ra^{+}}$. Our study presents an alternative approach for cavity-based quantum networks that is less sensitive to birefringent effects, and is applicable to a variety of atomic and solid-state emitter-cavity interfaces.
\end{abstract}

\maketitle

\section{Introduction}

Atom- and ion-based quantum processors promise opportunities for large-scale, high-fidelity quantum computation, enhanced quantum sensing and metrology, quantum simulation of many-body physics and quantum chemistry problems~\cite{decross2024computational,bluvstein2024logical,young2020half,franke2023quantum,guo2023site,ebadi2021quantum,hempel2018quantum}. However, scaling beyond thousands of qubits within a single processing unit (node) poses significant technical challenges~\cite{manetsch2024tweezer,malinowski2023wire}. This limitation has driven the development of quantum networks~\cite{kimble2008quantum, wehner2018quantum}, that utilize photons to connect distant physical systems and enable distributed quantum computation~\cite{cirac1999distributed,jiang2007distributed,main2025distributed}. For example, remote entanglement can be probabilistically established
between distant nodes via quantum teleportation which relies on interference of two photons that are entangled 
with stationary atomic qubits in each  node.~\cite{bouwmeester1997experimental,  duan2003efficient}. This approach overcomes some of the physical constraints of single-node processors and allows photonic-interconnected quantum systems to solve larger computational tasks~\cite{monroe2014large,li2024high,sinclair2024fault}.

The main challenge for remote entanglement generation (REG) is to improve the photon extraction efficiency while retaining high fidelity. In traditional free-space setups, the collection efficiency of spontaneously emitted photons is limited by the solid angle of the objective lenses~\cite{stephenson2020high, o2024fast,van2022entangling}. In contrast, optical cavities offer the potential for higher collection efficiency by modifying the atomic spontaneous emission, such that photons preferentially decay into the cavity mode~\cite{reiserer2015cavity,hartung2024quantum,schupp2021interface}. Despite their potential for enhanced efficiency, cavity-based quantum networks are particularly susceptible to errors arising from photon distinguishability due to a number of factors, including spontaneous emission, cavity jittering, and cavity birefringence~\cite{krutyanskiy2023entanglement}. Notably, cavity birefringence presents a substantial challenge for the commonly used polarization encoding scheme. In particular, micro-cavity systems~\cite{hunger2010fiber} are highly desirable because of the increased atom-cavity coupling~\cite{gehr2010cavity,takahashi2020strong} but they are more prone to birefringence. Manufacturing imperfections and mechanical stress give rise to ellipticity of the cavity mirrors~\cite{uphoff2015frequency,takahashi2014novel} that splits the energies of the two polarization eigenmodes, causing a time-dependent oscillation of the polarization states.
Furthermore, in long-distance
fiber links, polarization drift and fluctuations introduce additional
technical challenges for reliable polarization qubit measurements~\cite{van2022entangling,kucera2024demonstration}, which is another undesirable feature of polarization encoding for robust quantum networks. The inherent
susceptibility to birefringence together with the instability of polarization qubit measurements has stimulated the
search for alternative encoding methods, including the time-bin scheme, which has already shown promise in terms of 
higher fidelities, easy implementation of  quantum frequency
conversion, and its insensitivity to polarization errors, albeit
with typically lower entangling rates~\cite{knaut2024entanglement, saha2024high,uysal2024spin,li2025parallelized}. 



We propose an alternative approach to a cavity-based quantum
network that utilizes frequency encoding of photonic qubits~\cite{maunz2009heralded,connell2021ion}. This approach relies on matching the frequency difference
between two selected atomic qubit transitions to an even
multiple of the cavity’s free spectral range (FSR). In this way, both frequency qubit states can share the same polarization and resonate with
the fundamental mode of the cavity. Furthermore, the atom
can be located at the antinode position of the cavity center for both encoded frequencies with a maximal coupling strength.  

Our analytical and numerical analysis confirms that frequency encoding can achieve a high-degree of indistinguishability between the photonic wavepackets that enhances two-photon interference visibility, and is expected to improve the fidelity of REG in the presence of birefringence. The residual errors arise primarily from the difference in birefringence of the two networking nodes considered and can be further reduced by tuning other cavity parameters to improve the temporal mode matching. Finally, we outline and simulate two feasible experimental implementations with realistic physical parameters: one using fine structure splitting in the metastable D state of $\mathrm{^{40}Ca^+}$ and another using hyperfine splitting in the ground state of $\mathrm{^{225}Ra^+}$, indicating the practical use of our scheme.




The paper is organized as follows: in Section~\ref{sec:2}, we discuss the theoretical framework of atom-cavity quantum networks, including the use of two-photon interference for heralded entanglement and the sources of fidelity loss. We then introduce the frequency encoding scheme, detailing how it mitigates the errors from birefringence. Section~\ref{sec:3} focuses on the analytical modeling and derivation of the methods used in our simulations. Section~\ref{sec:4} shows the numerical results, including the analysis of the photonic wavepacket, two-photon interference visibility, and the robustness of entanglement fidelity under various cavity parameters.  In Section~\ref{sec:5}, we explore the experimental feasibility of implementing frequency encoding with different atomic species. Section~\ref{sec:6} concludes with a summary and an outlook for future experimental realization.

\section{Frequency encoding in atom-cavity systems}\label{sec:2}

In this section, we describe the basics of heralded entanglement in the context of performing two-photon interference. We also introduce the proposed frequency encoding scheme and discuss a few sources of fidelity loss.

\begin{figure}[!t]
    \begin{center}
        \includegraphics[width=0.5\textwidth]{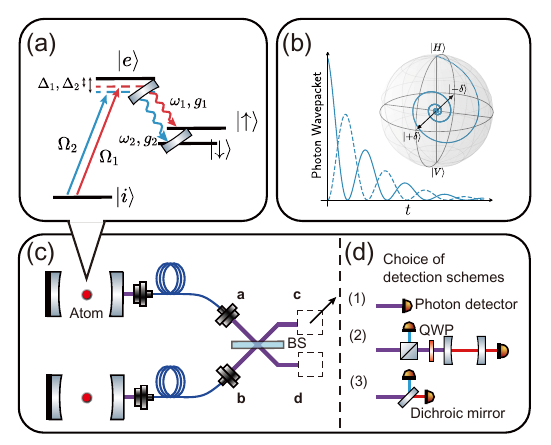}
        \caption{
        Frequency-encoded quantum network. (a) Atomic energy level diagram. The atom is first prepared in the initial state $\ket{i}$ and undergoes two simultaneous cavity-assisted Raman transitions (red and blue). Population is coherently transferred from  $\ket{i}$ to  $\ket{\uparrow}$ and $\ket{\downarrow}$ via virtual coupling to an excited state $\ket{e}$, associated with photons of two distinct colors emitted into two cavity modes.  Straight arrows denote the off-resonant laser drives and wavy arrows are photons mediated by atom-cavity coupling. This process entangles the atom's qubit state with the photonic qubit as $\frac{1}{\sqrt{2}}\left(\ket{\uparrow}\ket{{r}} + e^{i\theta}\ket{\downarrow}\ket{{b}}\right)$. (b) Polarization oscillation in a birefringent cavity. The solid ($\ket{H}$) and dashed ($\ket{V}$)  curves represent the temporal probability density of photon state decomposition in the two polarization basis. The state evolution is also plotted on a Poincar\'e sphere, with the amplitude representing the probability to detect a photon at time $t$.  (c) Photons from both nodes described in (a) are injected into a beam splitter to generate entanglement. Multiple measurement schemes of frequency-encoded photons can be applied at the output ports: (1) direct photon detection, (2) separation by a cavity with detection for each color, (3) separation by a dichroic mirror with detection for each color. The methods vary in efficiency and fidelity, allowing a  trade-off between performance and experimental simplicity.}
        \label{fig:1}
    \end{center}
\end{figure}

%

\subsection{Heralded entanglement based on two photon interference }

We consider an atom-cavity system with energy levels shown in Fig.\,\ref{fig:1}(a). At the beginning of each experimental cycle, the atom is initialized in the state $\ket{i}$. Then it simultaneously 
undergoes two cavity-assisted Raman transitions (CARTs)~\cite{kuhn2002deterministic,stute2012tunable} from $\ket{i}$ to qubit states $\ket{\uparrow}$ and $\ket{\downarrow}$, mediated by the excited state $\ket{e}$. Each transition is driven by an off-resonant laser (blue and red straight arrows), with the emitted photons coupled into cavity modes (blue and red curved arrows). The two resulting photonic states are required to have orthogonal characteristics, such as different polarizations, frequencies, or emission times, corresponding to different photonic qubit encoding schemes. If the strengths of both Raman transitions are tuned to be identical, the atom-photon system ends up in a maximally entangled state $\ket{\phi} = \frac{1}{\sqrt{2}}\left(\ket{\uparrow}\ket{0} + e^{i\theta}\ket{\downarrow}\ket{1}\right)$, where the $\ket{0}, \ket{1}$ basis represents a general photonic qubit. Depending on the chosen encoding scheme, the basis can be defined as $\ket{H}, \ket{V}$ (for horizontally and vertically polarized photons); $\ket{r}, \ket{b}$ (for photons with lower and higher frequency); $\ket{\text{early}},\ket{\text{late}}$ (for early or late emission), etc. The relative phase $
\theta$ is determined by the relative laser phase in the bichromatic drive. Without loss of generality, we will default to frequency encoding, and denote the photonic qubit as $\ket{r} \text{and} \ket{b}$ for the low- and high-frequency photon state, respectively.

After generating an atom-photon entangled state at each node, the two photons are sent to the input ports \textbf{a} and \textbf{b} of a beam splitter for two-photon interference~\cite{hong1987measurement,legero2004quantum}. The annihilation operators from each node $a$, $b$ are transformed by the beam splitter according to the unitary operation
\begin{equation}\label{bsop}
\begin{cases}
a = \frac{1}{\sqrt{2}}c-\frac{1}{\sqrt{2}}d,\\
b = \frac{1}{\sqrt{2}}c+\frac{1}{\sqrt{2}}d.
\end{cases}
\end{equation}

Measurements are then performed at output ports \textbf{c} and \textbf{d} in the $\ket{r}, \ket{b}$ basis (Fig.\,\ref{fig:1}(c)). Ideally, with photons from both nodes being perfectly indistinguishable, the state after the beam splitter transformation is given by:

\begin{align}
\ket{\Phi} =& \frac{1}{4} \Bigg[ \sqrt{2}\Big( \ket{{r}_c {r}_c} - \ket{{r}_d {r}_d} \Big) \ket{\uparrow \uparrow}\notag\\
&+ \sqrt{2}\Big(\ket{{b}_c {b}_c} - \ket{{b}_d {b}_d} \Big) \ket{\downarrow \downarrow} \notag\\
&+ \ket{{r}_c {b}_c} \Big( \ket{\uparrow \downarrow} + \ket{\downarrow \uparrow} \Big) 
- \ket{{r}_d {b}_d} \Big( \ket{\uparrow \downarrow} + \ket{\downarrow \uparrow} \Big) \notag\\
&+ \ket{{r}_c {b}_d} \Big( \ket{\uparrow \downarrow} - \ket{\downarrow \uparrow} \Big) 
- \ket{{r}_d {b}_c} \Big( \ket{\uparrow \downarrow} - \ket{\downarrow \uparrow} \Big) \label{eqn:state_after_BS}
\Bigg],
\end{align}
where the photonic qubit is denoted such that $\ket{{r}_c {r}_c} = c^\dagger_r c^\dagger_r \ket{0}$ represents the state causing two red photons to arrive at port \textbf{c}. The equation shows a combination of all possible coincidence events and the corresponding atom-atom state after photon detection. For example, a successful detection of one red photon in port \textbf{c} and one blue photon in port \textbf{d} heralds the asymmetric Bell state $\ket{\psi^{-}} = \left(\ket{\uparrow\downarrow}-\ket{\downarrow\uparrow}\right)/\sqrt{2}$. The generation of such a maximally entangled state relies on the indistinguishability of photons from the two nodes: when we detect $\ket{{r}_c{b}_d}$ and the path information is erased, the atom-atom state collapses to a superposition of $\ket{\downarrow\uparrow}$ and $\ket{\uparrow\downarrow}$ .

We note that in order to herald on both asymmetric Bell states $\ket{\Psi^\pm}$, the realization of the measurement depends on the chosen encoding scheme. For polarization-encoded photons, polarization beam splitters (PBS) can be used at output ports \textbf{c} and \textbf{d}, with two photon detectors after the output ports of each PBS. For frequency-encoded photons, a second cavity resonating with only one of the two frequencies can distinguish the frequency qubit, as shown in Fig.\,\ref{fig:1}(d)(2). Also, a dichroic mirror (3) can be applied if the frequency difference is beyond a few THz. Directly detecting the photons that appears at both ports (1) without distinguishing their frequencies is also feasible~\cite{maunz2009heralded}. This approach simplifies detection, but cannot herald $\ket{\Psi^+}$, reducing efficiency by 50\% and possibly leading to additional fidelity loss with nonidentical photons due to anti-bunching effects~\cite{stephenson2020entanglement}. 

\subsection{Sources of fidelity loss}\label{sec:2.2}

While Eq.~\ref{eqn:state_after_BS} presents an ideal atom-atom entanglement result, in practice multiple factors can lead to imperfect fidelity. 
For instance, a difference in photonic wavepacket shape between the two nodes reduces the final fidelity because it introduces distinguishability. Such a difference could arise from the different parameter settings between nodes. The distinguishability of the two nodes from  probabilistic events such as re-excitation errors (discussed in detail in Section \ref{sec:3.4}), and technical noise such as cavity jitter \cite{krutyanskiy2023entanglement}, also contribute to reduced fidelity.

Mirror-induced birefringence is another important source of fidelity loss. Although cavity-based photon collection schemes may enhance collection efficiency, their mirrors often exhibit birefringence because of inherent ellipticity and mechanical stress, especially in microcavities. The mirror birefringence introduces an energy splitting to the non-degenerate modes of polarization. As a result, superpositions of the polarization eigen-states of photons in the cavity rotate as a function of time. Fig.\,\ref{fig:1}(b) illustrates the evolution of a $\ket{H}$ photon emitted in a birefringent cavity with coupling parameters of $(\kappa,\gamma,\delta) = (0.01,\,0,\,0.05)\,g$. Spontaneous atomic decay $\gamma$ is ignored and a small cavity decay $\kappa$ is chosen here to illustrate the polarization oscillation in the cavity. Here, $g$ represents the atom-cavity coupling rate and $2\delta$ is the birefringent splitting. The cavity eigenmodes $\ket{\pm\delta}$ are set to  $\left(\ket{H}\pm\ket{V}\right)/\sqrt{2}$, causing the $\ket{H}$ photon to undergo a rotation while its amplitude gradually decreases. The left side shows the oscillations in projection onto the horizontal and vertical polarization basis, while the right side shows the corresponding trajectory on the Poincar\'e sphere.  This effect causes the entangled atom-photon state to be time-dependent. Consequently, the heralded atom-atom state will include extra unwanted components in $\ket{\uparrow\uparrow}$ and $\ket{\downarrow\downarrow}$ that depend on the detection, and more importantly, on the unknown emission times. 


To address the birefringence effect, in addition to improving mirror manufacturing~\cite{takahashi2014novel,garcia2018dual}, some pre- and post-herald correction methods have been proposed to compensate for such rotation by applying a time-dependent unitary transformation to the photonic or atomic qubit ~\cite{kassa2023effects}. Here we propose frequency encoding as a robust alternative to mitigate the effect of birefringence on the fidelity. Although birefringence still induces polarization rotation, photons encoded with distinct frequencies can be distinguished during measurement. Thus, even if the photon has a time-varying polarization in the cavity, the frequency-dependent heralding procedure yields results that are uniquely correlated to the atom's decay path. With no misidentifications caused by the birefringence-induced polarization rotation, the error rate can be substantially reduced. While fidelity loss due to the distinguishability in the polarization degree-of-freedom may still be present, this effect may be suppressed by choosing the same polarization for both emission paths.

\section{Models and Methods}\label{sec:3}
This section provides a detailed derivation of the REG mechanism, with the fidelity-reducing error sources discussed in Section \ref{sec:2.2} taken into account. The equations provided here will be used for the numerical simulations discussed in Section \ref{sec:4}. In Section \ref{sec:3.1} we discuss the atom-photon entanglement generation process, focusing on the system's pure and mixed evolution based on its Hamiltonian.  Section \ref{sec:3.2} and \ref{sec:3.3} present the two-photon interference process and the resulting REG fidelity.

\subsection{Atom-photon entanglement} \label{sec:3.1}
During the generation of entangled atom-photon pairs, each node operates independently and simultaneously. We consider an atom going through two simultaneous cavity-assisted Raman transitions (CARTs), as shown in Fig.\,\ref{fig:1}(a). Four levels are involved in the transition, where $\ket{i}$ is the initial state, $\ket{e}$ is the excited state and $\ket{\uparrow}$ and $\ket{\downarrow}$ are the qubit states. During each REG attempt, the atom is driven from $\ket{i}$ to qubit states $\ket{\uparrow}$ and $\ket{\downarrow}$ with equal strength, through an off-resonant bichromatic laser coupling $\ket{i}$ and $\ket{e}$, accompanied by emission of cavity photons. In frequency encoding, the emitted photons should share the same polarization to ensure that they evolve identically under birefringence. Without loss of generality, we assume that the photons are horizontally polarized.

The Hilbert space of the atom-cavity system is described by the tensor product of the atomic and photonic state. Here, only the states involved in the Raman process are considered: $\ket{i}$, $\ket{e}$, $\ket{\uparrow}\ket{rH}$, $\ket{\uparrow}\ket{rV}$, $\ket{\downarrow}\ket{bH}$,  $\ket{\downarrow}\ket{bV}$, where $\ket{r}$ and $\ket{b}$ denote the red and blue photonic frequency states, and $\ket{H}$ and $\ket{V}$ represent the horizontal and vertical polarization. Evolution of the system  in the CART follows the Lindblad equation:
\begin{equation}
    \dot{\rho} = -\frac{i}{\hbar}\left(H\rho-\rho H^\dagger \right)+\sum_j  \left( L_j \rho L_j^\dagger - \frac{1}{2}\{L_j^\dagger L_j, \rho\}\right),\label{eqn:mastereqn}
\end{equation}

where the full Hamiltonian can be written as:

\begin{equation}
    H = H_0 + H_{\text{L}} + H_{\text{Cav}} + H_{\text{Bire}} + H_{\kappa} + H_{\gamma}.
\end{equation}

Here, $H_0$ represents the bare atomic energy levels, $H_{\text{L}}$ denotes the atom-laser interaction and $H_{\text{Cav}}$ denotes the atom-cavity interaction. The birefringence of the cavity mirrors is modeled using the term $H_{\text{Bire}}$, which introduces an energy splitting of $2\hbar\delta$ to the polarization modes. $H_{\kappa}$ and $H_{\gamma}$ are the non-Hermitian parts of the Hamiltonian.   The term $H_{\kappa}$ describes the dynamics of photons leaking out of the cavity, and can be described with collapse operators in the Lindblad treatment, e.g. $L_{rH}=\sqrt{2\kappa}\ket{\uparrow,0}\bra{\uparrow,rH}$ for a red and horizontally polarized photon. (In the following discussion, $\ket{0}$ represents the vacuum state in which no cavity photon is present in any mode.) Since the two projected states $\ket{\uparrow,0}$ and $\ket{\downarrow,0}$ are not included in the Hilbert space, this term can be treated as a non-hermitian term that represents population leakage from the metastable states $\ket{\uparrow}$ and $\ket{\downarrow}$. Similarly, $H_{\gamma}=-i\gamma_{xe}\ket{e}\bra{e}$ describes a spontaneous emission from $\ket{e}$ to any other level $\ket{x}$ not resonant with the laser drives in the CART, including $\ket{\uparrow}$, $\ket{\downarrow}$ and other levels not explicitly considered. It is an effective description of the collapse operator $L_{xe} = \sqrt{2\gamma_{xe}}\ket{x}\bra{e}$. Thus, the Hamiltonian has the form
\begin{widetext}
\begin{align}
&\;\;\;\quad\quad\quad\quad\ket{i}\;\quad\quad\quad\ket{e}\;\;\quad\quad\ket{\uparrow,{rH}}\quad\;\;\ket{\uparrow,{rV}}\quad\quad\quad\ket{\downarrow,{bH}}\quad\quad\quad\quad\;\;\ket{\downarrow,{bV}}\notag\\
& H/\hbar=\begin{pmatrix} 
\delta_{\text{Stark}} & \frac{1}{2}\Omega(t) \\  
\frac{1}{2}\Omega^*(t) & -\Delta_{1}-i\gamma_{xe} & g_{1} & & g_{2} \\  
& g_{1} & \delta_{z}-i\kappa & \delta_{x}-i\delta_{y} \\  
& & \delta_{x}+i\delta{y} & -\delta{z}-i\kappa\\  
& g_{2} & & & \Delta_{2}-\Delta_{1}+\delta_{z}-i\kappa & \delta_{x}-i\delta_{y} \\  
& & & & \delta_{x}+i\delta_{y} & \Delta_{2}-\Delta_{1}- \delta_{z}-i\kappa
\end{pmatrix},\label{eqn:Hamiltonian}
\end{align}
\end{widetext}
where we denote the combined bichromatic laser drives as $\Omega(t) = \Omega_1 + \Omega_2 e^{i(\Delta_2-\Delta_1)t}$, which creates a Stark shift of $\delta_{\text{Stark}} = \Omega_1^2/(4\Delta_1)+\Omega_2^2/(4\Delta_2)$.
Here, $\Delta_{j}$, $\Omega_{j}$ and $g_{j}$ represent the laser detuning, drive Rabi frequency and atom-cavity coupling strength of the $j^{\text{th}}$ Raman transition ($j \in \{1, \ 2\}$).  The birefringence term $(\delta_x, \delta_y, \delta_z) = \delta\vec{e}_{\delta}$ are the components of the Stokes vector on the Poincar\'e sphere, with $\vec{e}_{\delta}$ describing the orientation of the two orthogonal cavity eigenmodes~\cite{barrett2019polarization}, and the magnitude $\delta$ being half the frequency splitting between them.

The collapse operators $L_j$ used in the Lindblad equation is:
\begin{equation}
  L_{ie} = \sqrt{2\gamma_{ie}}\ket{i}\bra{e}.
\label{eqn:collapseoperators}
\end{equation}

$L_{ie}$ describes the spontaneous emission from $\ket{e}$ back to $\ket{i}$. This collapse event is specially treated because population driven off-resonantly from $\ket{i}$ to $\ket
{e}$ that leaks back to $\ket{i}$ will remain in the subspace addressed by the laser drives, possibly resulting in emission of a delayed photon. This re-excitation process contributes to part of the fidelity loss, and will be elaborated on later in this section. In contrast, collapse events from $L_{xe}$ do not result in the emission of cavity photons. Consequently, decay to these levels implies failure to generate a photon in the cavity mode. While such events reduce the REG rate, they do not directly affect the fidelity of heralded entanglement.

In order to find the atom-photon density matrix, we adopt the approach of \cite{krutyanskiy2023entanglement} as follows: the eventual detection of a photon necessary for a heralding event can only occur if a collapse event $L_{xe}$ has not occurred. Collapse events $L_{ie}$ lead to a mixing of the temporal Hilbert space of the emitted photon. As such, we first seek to describe the state $\ket{\phi}$ of a pure photon that is extracted from the cavity with no collapse events during its evolution. We call this the "pure branch". Later, in Section~\ref{sec:3.4}, we will explicitly consider the impact of collapse events $L_{ie}$ on the fidelity of the atom-atom entanglement scheme. The pure branch density matrix $\rho_0 = \ket{\phi}\bra{\phi}$ can be found from Eq.\,\ref{eqn:mastereqn} by removal of terms of the form $L_j \rho L_j^\dagger$:
\begin{align}
    \dot{\rho}_0 &= -\frac{i}{\hbar}\left(H\rho_0-\rho_0 H^\dagger \right)-\sum_j \frac{1}{2}\{L_j^\dagger L_j, \rho_0\} \notag \\&= -\frac{i}{\hbar} \left(H_{\text{eff}}\rho_0-\rho_0 H_{\text{eff}}^\dagger \right),\label{eqn:mastereqn_noiseless}
\end{align}
where \begin{equation}
    H_{\text{eff}} = H -\sum_j\frac{i}{2}L_j^\dagger L_j.
\end{equation} 
In this case we can calculate the pure evolution of the photon by applying the Schr\"odinger equation to our non-Hermitian effective Hamiltonian:
\begin{align}
    \dot{\ket{\phi}} &= -\frac{i}{\hbar}H_{\text{eff}}\ket{\phi}.\label{eqn:schrodinger}
\end{align}
The leakage of photon can be described using creation operators of cavity photons and leaked photons  defined by $a^\dagger(t) = \sqrt{2\kappa}a_{\text{cav}}^\dagger(t)$~\cite{goto2019figure}.

Now we consider the output state of the photon from the cavity. For notational brevity, we make the choice to consider a red and horizontally polarized photon. The wave function of the output photon is
\begin{equation}
\ket{\phi^\text{out}_{rH}} = \sqrt{2 \kappa}\int \braket{\uparrow,rH|\phi(t)}a^{\dagger}_{rH}(t)\ket{0}dt.
\end{equation}
For simplicity we write it as 
\begin{equation}
    \ket{\phi^\text{out}_{rH}} = \int \varphi_{rH}(t)a^{\dagger}_{rH}\ket{0}dt=\tilde{a}_{rH}^\dagger\ket{0},
\end{equation}
where $\varphi_{rH}(t) = \sqrt{2 \kappa}\braket{\uparrow,rH|\phi(t)}$. The derivation is analogous to that for other frequencies and polarizations. Finally, the atom-photon state can be written as:
\begin{align}
\ket{\phi}_{\text{a-p}} &= \ket{\uparrow}\left(\tilde{a}_{rH}^\dagger+\tilde{a}_{rV}^\dagger\right)\ket{0}
+\ket{\downarrow}\left(\tilde{a}_{bH}^\dagger+\tilde{a}_{bV}^\dagger\right)\ket{0} \label{eqn:atom-photon},
\end{align}
which exhibits entanglement between the atomic qubit $\ket{\uparrow}, \ket{\downarrow}$ and the photonic qubit $\ket{r}, \ket{b}$.

\subsection{Atom-atom entanglement} \label{sec:3.2}
As shown in Fig.\,\ref{fig:1}\,(c), after the entangled atom-photon pairs are generated, the photons that leak out of the cavities are guided into optical fibers to the two input ports of a 50:50 beam splitter, and the photonic states are measured after each output port.

The full two-node quantum state immediately before interference on the beam splitter can be written as the tensor product of two single-node systems, as described in Eq.\,\ref{eqn:atom-photon}:
\begin{align}
\ket{\Phi} &= \left(\ket{\uparrow}\left(\tilde{a}_{ArH}^\dagger+\tilde{a}_{ArV}^\dagger\right)
+\ket{\downarrow}\left(\tilde{a}_{AbH}^\dagger+\tilde{a}_{AbV}^\dagger\right)\right) \notag\\
&\otimes \left(\ket{\uparrow}\left(\tilde{b}_{BrH}^\dagger+\tilde{b}_{BrV}^\dagger\right)
+\ket{\downarrow}\left(\tilde{b}_{BbH}^\dagger+\tilde{b}_{BbV}^\dagger\right)\right)\ket{0},
\end{align}
where the subscripts $A$ and $B$ refer to nodes A and B. We then apply the beam splitter transformation in Eq. \ref{bsop}, which gives the following output state:
\begin{align}
\ket{\Phi} &= \left(\ket{\uparrow}\left(\tilde{c}_{ArH}^\dagger-\tilde{d}_{ArH}^\dagger+\tilde{c}_{ArV}^\dagger-\tilde{d}_{ArV}^\dagger\right)\right.\notag\\
&\left.+\ket{\downarrow}\left(\tilde{c}_{AbH}^\dagger-\tilde{d}_{AbH}^\dagger+\tilde{c}_{AbV}^\dagger-\tilde{d}_{AbV}^\dagger\right)\right)\notag\\
&\otimes \left(\ket{\uparrow}\left(\tilde{c}_{BrH}^\dagger+\tilde{d}_{BrH}^\dagger+\tilde{c}_{BrV}^\dagger+\tilde{d}_{BrV}^\dagger\right)\right.\notag\\
&\left.+\ket{\downarrow}\left(\tilde{c}_{BbH}^\dagger+\tilde{d}_{BbH}^\dagger+\tilde{c}_{BbV}^\dagger+\tilde{d}_{BbV}^\dagger\right)\right)\ket{0}.
\end{align}
Here, we again drop the explicit time dependence notation for simplicity. For example, the creation operators in the output ports are given by  $\tilde{c}_{ArH}^\dagger = \int \varphi_{ArH}(t) c_{rH}^\dagger(t) dt$, for a red and horizontally polarized photon generated from node A.

We consider coincidence events with one red and one blue photon each detected in port \textbf{c} using detection scheme (3) that ideally herald the atom-atom state $\ket{\Psi^{+}} = \left(\ket{\uparrow\downarrow}+\ket{\downarrow\uparrow}\right)/\sqrt{2}$. By applying its corresponding projection operator
\begin{align}
    P_{cc}^\text{freq}(t_r,t_b)&=\sum_{x,y=H,V}c_{rx}(t_r)^\dagger c_{by}(t_b)^\dagger \ket{0}\bra{0} c_{rx}(t_r)c_{by}(t_b)
\end{align}
the projected state is given by
\begin{align}
\ket{\Phi_{cc}^\text{freq}}&=
P_{cc}^\text{freq}\ket{\Phi}\notag \\
&=\sum_{x,y =H,V} \bigg[
\ket{\uparrow \downarrow} \varphi_{Arx}(t_r) \varphi_{Bby}(t_b)\notag\\
&\quad+ \ket{\downarrow \uparrow} \varphi_{Aby}(t_b) \varphi_{Brx}(t_r) \bigg]c^\dagger_{rx}(t_r) c^\dagger_{by}(t_b)\ket{0}.
\label{eqn:project}
\end{align}


After tracing out the photon states from Eq.~\ref{eqn:project}, the atom-atom state takes the following form:
\begin{equation}
    \rho_{cc}^\text{freq}(t_r,t_b) = \text{Tr}_p \ket{\Phi_{cc}^\text{freq}}\bra{\Phi_{cc}^\text{freq}}
=\sum_{x,y=H,V}\ket{\Psi_{xy}}\bra{\Psi_{xy}},\label{eqn:freq_entangle}
\end{equation}
where
\begin{equation}
   \ket{\Psi_{xy}} = \ket{\uparrow\downarrow}\varphi_{Arx}(t_r)\varphi_{Bby}(t_b) + \ket{\downarrow\uparrow}\varphi_{Aby}(t_b)\varphi_{Brx}(t_r)  .\label{eqn:freq_entangle2}
\end{equation}


Similarly, we derive the atom-atom entangled state for polarization encoding:
\begin{align}
&\ket{\Psi_{cc}^\text{pol}(t_H,t_V)}\notag\\
&=\ket{\uparrow\uparrow}\Big(\varphi_{AHH}(t_H)\varphi_{BHV}(t_V)+\varphi_{AHV}(t_V)\varphi_{BHH}(t_H)\Big)\notag\\
&+\ket{\uparrow\downarrow}\Big(\varphi_{AHH}(t_H)\varphi_{BVV}(t_V)+\varphi_{AHV}(t_V)\varphi_{BVH}(t_H)\Big)\notag\\
&+\ket{\downarrow\uparrow}\Big(\varphi_{AVH}(t_H)\varphi_{BHV}(t_V)+\varphi_{AVV}(t_V)\varphi_{BHH}(t_H)\Big)\notag\\
&+\ket{\downarrow\downarrow}\Big(\varphi_{AVH}(t_H)\varphi_{BVV}(t_V)+\varphi_{AVV}(t_V)\varphi_{BVH}(t_H)\Big),\label{eqn:pol_entangle}
\end{align}
where the indices are arranged such that $AHV$ stands for a photon from node $A$ with initial horizontal polarization and rotated to vertical under birefringence. While the measurement is performed on the actual polarization, represented by the third index, the second index is still tracked because it indicates the ion state to which the photon is entangled. To understand Eq. \ref{eqn:pol_entangle}, we examine the first term proportional to $\ket{\uparrow \uparrow}$. This term shows that if the photon from node $B$ ($A$) was initially horizontal and was rotated to vertical under birefringence, this photon can be detected along with one horizontal photon from node $A$ ($B$) that was not rotated under birefringence, and still register as a coincidence event with both detections on path $c$. 
Similarly, a $V\rightarrow H$ rotation from either node may introduce an admixture of $\ket{\downarrow \downarrow}.$ This reduces the overlap with the desired asymmetric Bell state.

By comparing this to Eq.~\ref{eqn:freq_entangle2}, we can see the advantage of frequency encoding. In the ideal case where wavefunctions from both nodes for the same frequency are identical (e.g., $\varphi_{Arx}(t_r) = \varphi_{Brx}(t_r) = \varphi_{rx}(t_r)$), the frequency encoded entangled state in Eq.~\ref{eqn:freq_entangle} can be simplified to $\ket{\Psi_{xy}} = \varphi_{rx}(t_r)\varphi_{by}(t_b)\left(\ket{\uparrow\downarrow} + \ket{\downarrow\uparrow}\right)$, corresponding to an ideal Bell state with unity fidelity. In polarization encoding, the heralded state is still not ideal under this assumption, due to the aforementioned admixtures of $\ket{\uparrow\uparrow}$ and $\ket{\downarrow\downarrow}$ in Eq.~\ref{eqn:pol_entangle}. In the no-birefringence limit, $\varphi_{HV}(t) = \varphi_{VH}(t) = 0$, the atom-atom entangled state is again proportional to $\ket{\uparrow \downarrow} + \ket{\downarrow \uparrow}$, as in the case for frequency encoding.

\begin{figure}[b]
    \centering
    \includegraphics[width=0.48\textwidth]{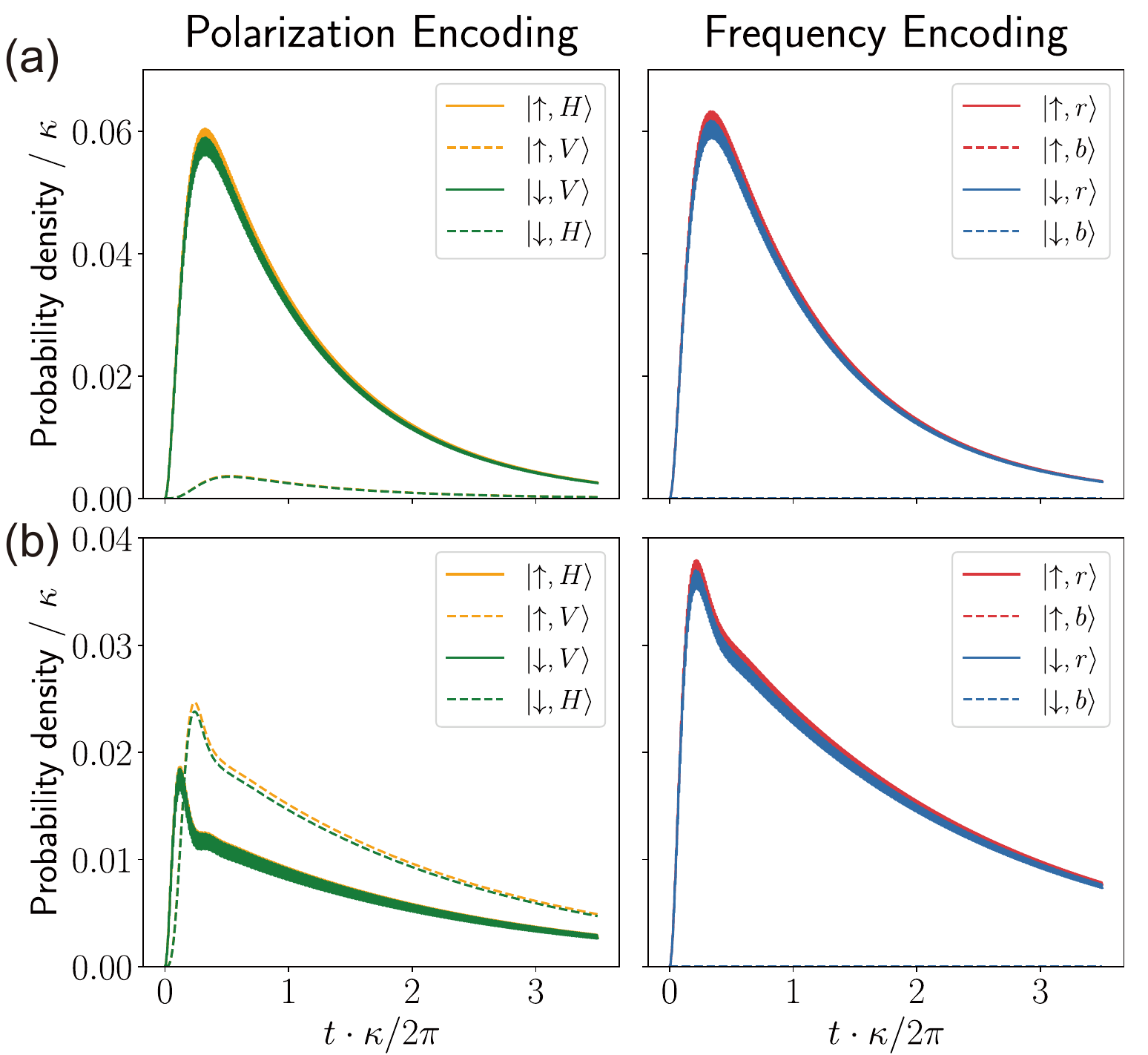}
    \caption{Probability density of atom-photon correlation outcomes based on the polarization and frequency encoding schemes. (a) and (b) compare the wavepackets under different birefringence magnitudes: (a) $\delta = 0.3\kappa$ and (b) $\delta = 2\kappa$. Photons encoded by polarization (left) are represented by orange and green lines, while photons encoded by frequency (right) are represented by blue and red lines.
    The solid lines denote the probability density associated with the  expected  entangled atom-photon correlations, whereas the dashed lines denote the probability density of the incorrect correlations due to birefringence. The polarization degree of freedom is traced out for the frequency encoded case.}
    \label{fig:2}
\end{figure}

\begin{figure*}[t]
    \centering
    \includegraphics[width=1\textwidth]{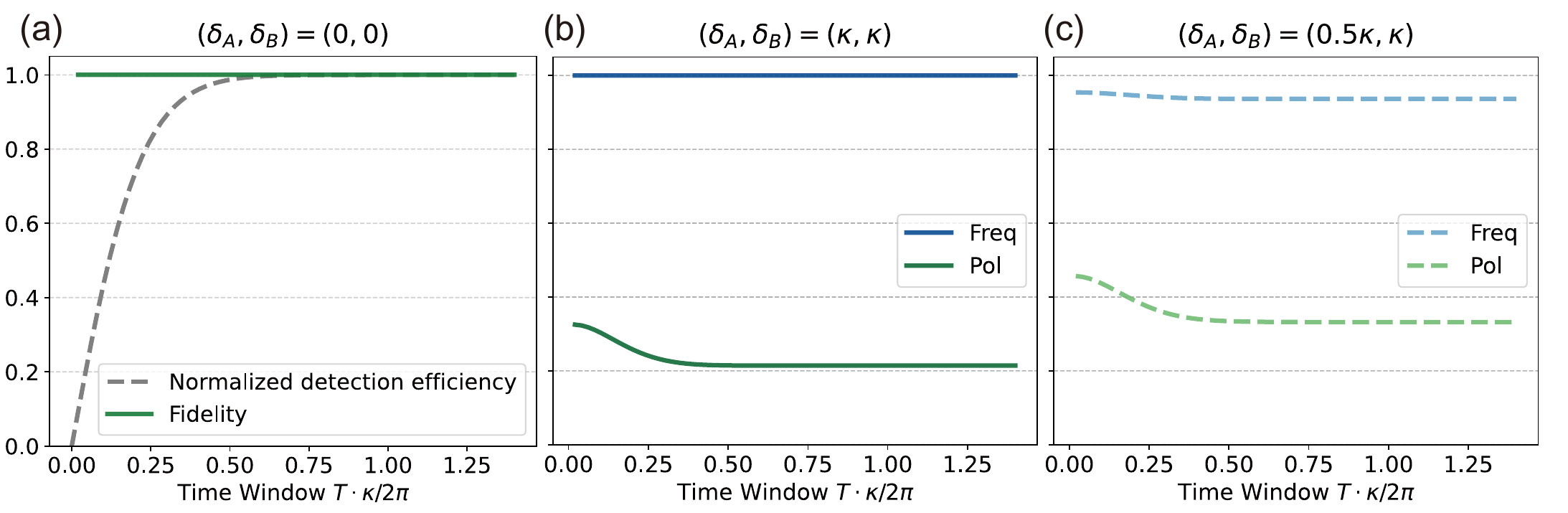}
    \caption{
    Detection efficiency and atom-atom fidelity as a function of coincidence window length under different encoding schemes. 
    The same cavity parameters are used for both frequency and polarization encoding schemes. (a) Detection efficiency and fidelity in the absence of birefringence. Both encoding schemes give unity fidelity. (b) and (c): Atom-atom fidelity under the same ((b)) and  different ((c)) birefringence magnitude between the two nodes with frequency and polarization encoding.}
    \label{fig:3}
\end{figure*}

\subsection{Entanglement Fidelity} \label{sec:3.3}

Using the atom-atom entangled state derived from Eq.\,\ref{eqn:freq_entangle}, the probability distribution of coincidence events where the photon pairs are detected at times $\left(t_r,t_b\right)$, can be calculated from the norm of the projected state:
\begin{align}
p(t_r,t_b)=&\sum_{x,y = H,V} \left|\varphi_{Arx}(t_r)\varphi_{Bby}(t_b)\right|^2\notag\\& + \left|\varphi_{Aby}(t_b)\varphi_{Brx}(t_r)\right|^2\:.
\label{eqn:p}
\end{align}
The corresponding atom-atom entanglement fidelity to the asymmetric Bell state is:
\begin{align}
F(t_r,t_b) &= \frac{1}{2p(t_r,t_b)} \sum_{x,y=H,V} \Big( \left|\varphi_{Arx}(t_r)\varphi_{Bby}(t_b)\right| \notag\\
&\quad + \left|\varphi_{Aby}(t_b)\varphi_{Brx}(t_r)\right| \Big)^2,
\label{eqn:F}
\end{align}
where we have chosen the relative phase of the Bell state to maximize the overlap with $\rho_{cc}^\text{freq}$.

Note that if we use the detection scheme (2) shown in Fig.\,\ref{fig:1}(c), photons with rotated polarization ($\ket{V}$ here) will be rejected by a PBS before being detected. In this case the summation in Eq.\,\ref{eqn:p} and Eq.\,\ref{eqn:F}  can be removed by taking $x,y$ as $H$. The fidelity will be slightly increased at the cost entanglement generation rate. In the remainder of this paper, we will use the detection scheme (3) as the default, where polarization is not filtered.

\begin{figure}[!t]
    \centering    \includegraphics[width=0.48\textwidth]{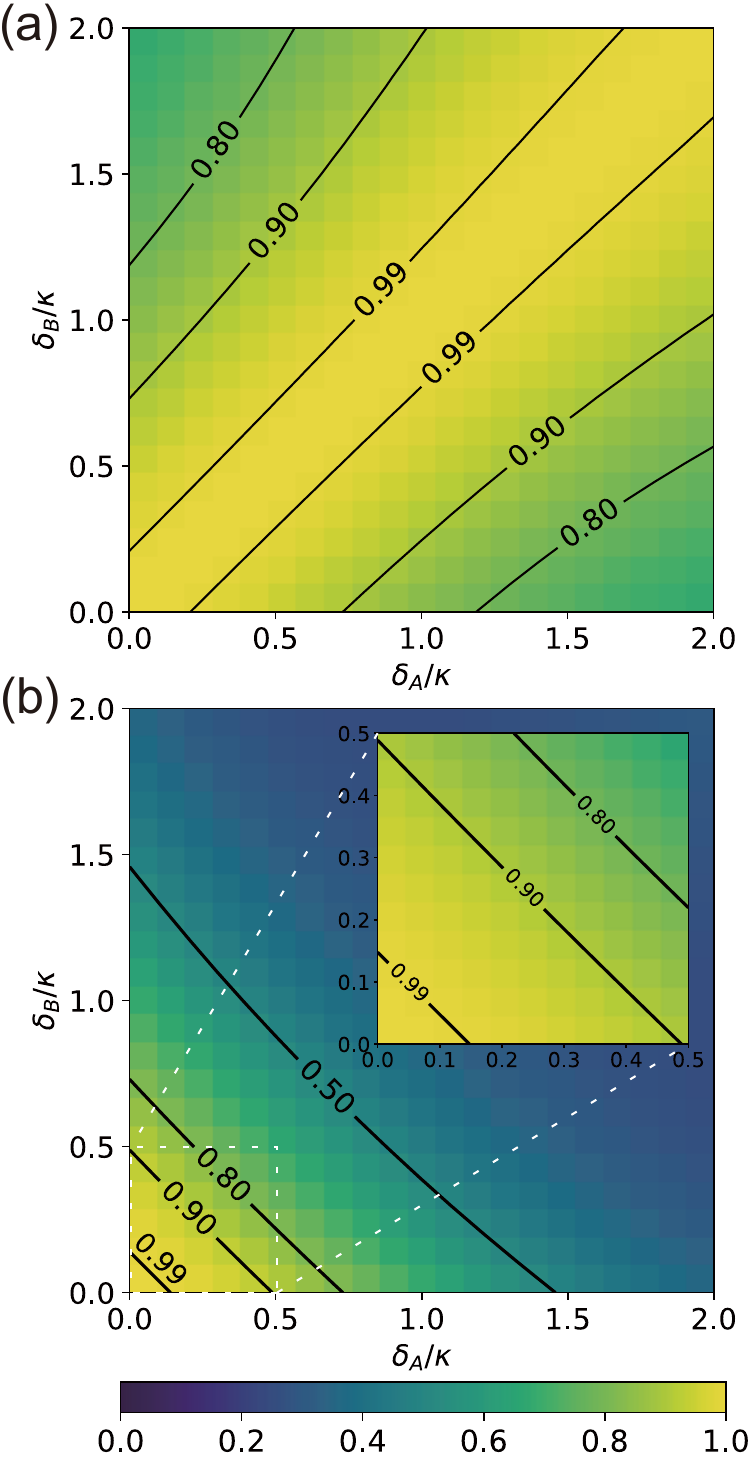}
    \caption{
    Atom-atom fidelity versus birefringence magnitude at each node, using (a) frequency encoding and (b) polarization encoding. To demonstrate the effect of birefringence on the fidelity, results are derived from simulations with otherwise identical parameters and no re-excitation.}
    \label{fig:4}
\end{figure}

\subsection{Re-excitation} \label{sec:3.4}
In Section \ref{sec:3.1} we derived the pure branch of the photon's evolution where no re-excitation events occur. However, Eq.~\ref{eqn:F} does not capture the effect of re-excitation errors from collapse events $L_{ie}$ on the temporal purity of the cavity-emitted photons.  
In this section we extend our model to account for re-excitation, as this process is one of the most detrimental fidelity-reducing effects for remote entanglement generation (REG) in many atom-cavity systems~\cite{krutyanskiy2023entanglement,walker2020improving, meraner2020indistinguishable}.

Re-excitation events affect the REG fidelity by introducing mixing of the photonic wavepacket in the temporal Hilbert space. In the transition scheme shown in Fig.\,\ref{fig:1}(a), the excited state $\ket{e}$ will typically have a lifetime on the order of nanoseconds, which is much shorter than the Raman pulse~\cite{walker2020improving}.
Decay from $\ket{e}$ to states other than the initial state $\ket{i}$ reduces the number of successful attempts, but because this process does not produce a photon in the cavity mode, it does not reduce the atom-atom entanglement fidelity. However, decay from $\ket{e}$ to $\ket{i}$ can still result in subsequent production of a photon in the cavity mode. This can be seen as an incoherent process that results in mixing of the photon's state in the temporal Hilbert space, diminishing the REG fidelity \cite{meraner2020indistinguishable,krutyanskiy2023entanglement}. Modification to the Raman drive pulse, such as increasing of the detuning and pulse shaping ~\cite{tanji2024rate}, and careful choice of atomic level scheme ~\cite{walker2020improving,kikura2024high} can mitigate the effects of re-excitation errors.

To extend our model to account for re-excitation errors, we consider the photon emission time $s$ to be a random variable with probability distribution $P(s)$. Then we may write the density matrix $\rho_\text{re-ex}$ as an integral of pure branch evolution density matrices $\ket{\phi(t-s)}\bra{\phi(t-s)}$, weighted by the probability distribution for an event $L_{ie}$ to occur:

\begin{equation}
    \rho_\text{re-ex}(t) = \int P(s) \ket{\phi(t-s)}\bra{\phi(t-s)}ds.
\end{equation}
Here, $\ket{\phi(t)}$ is the pure photon state from (\ref{eqn:schrodinger}), and $P(s)$ is given by
\begin{equation}
    P(s) = \boldsymbol{\delta}(s)+\bra{e}\rho(s)\ket{e},
\end{equation}
where the additional delta function $\boldsymbol{\delta}(s)$ centered at time $s=0$ accounts for the initial pure branch where no collapse events occur. $\rho(s)$ is the mixed branch density matrix from Eq.~\ref{eqn:mastereqn}. 

Similar to Eq.\,\ref{eqn:p}, we can derive the final detection probability:
\begin{align}
p(t_r,t_b)=&\int P_A(s_A)P_B(s_B)ds_A ds_B \times \notag \\
&\sum_{x,y = H,V} \left|\varphi_{Arx}(t_r-s_A)\varphi_{Bby}(t_b-s_B)\right|^2 \notag\\& + \left|\varphi_{Ary}(t_b-s_A)\varphi_{Brx}(t_r-s_B)\right|^2.
\end{align}
Finally, the atom-atom entanglement fidelity considering the re-excitation errors is given by
\begin{align}
F(t_r,t_b) =& \frac{1}{2p(t_r,t_b)} \int P_A(s_A)P_B(s_B)ds_A ds_B\times \notag\\ &\sum_{x,y=H,V} \Big( \left|\varphi_{Arx}(t_r-s_A)\varphi_{Bby}(t_b-s_B)\right| \notag\\&+ \left|\varphi_{Aby}(t_b-s_A)\varphi_{Brx}(t_r-s_B)\right| \Big)^2.
\end{align}

\section{Numerical results}\label{sec:4}

Due to the complexity of the Hamiltonian in Eq.~\ref{eqn:Hamiltonian}, an analytical solution is not feasible. We conducted further numerical simulation on the CART and heralded entanglement based on two-photon interference. In this section we provide an overview of our simulation based on the model described in Section \ref{sec:3}. The advantages of the frequency encoding scheme are demonstrated by comparing entanglement fidelity under frequency encoding with that of the traditional polarization encoding scheme. Our results show that frequency encoding is robust against birefringence, whereas entanglement fidelity in polarization encoding deteriorates significantly when birefringence reaches the same order as the cavity decay rate $\kappa$. 

For simplicity and to demonstrate the effect of birefringence, here we discard re-excitation events and assume that all other cavity parameters are identical for both nodes, so that any possible reduction in fidelity is caused by birefringence. The cavity eigenmodes are chosen as $\left(\ket{H}\pm\ket{V}\right)/\sqrt{2}$, corresponding to a Hamiltonian $H_{\text{Bire}}=\hbar\delta\sigma_x$. This choice maximizes rotation of the initial polarization, and results in the largest fidelity loss~\cite{kassa2023effects}. We first conduct simulations on the atom-photon output states. Fig.\,\ref{fig:2} shows some typical shape of photonic wavepackets generated from Hamiltonian in Eq.~\ref{eqn:Hamiltonian}. The plotted curves show the emission probability of different atom-photon states measured at the output port as a function of time. The wavepackets are extracted from the pure state density matrix, which evolves following the regular Lindblad equation (Eq.~\ref{eqn:mastereqn}). The Rabi frequencies are adjusted such that the photonic wavepackets of the two desired output states are well overlapped for maximal fidelity. Fig.\,\ref{fig:2}(a) shows the wavepackets under small birefringence $\delta = 0.3\kappa$. 
For polarization-encoded photons, a fraction of each branch is rotated to its orthogonal polarization state, and the wavepacket is also balanced between two rotated components. In the case of frequency encoding, although similar polarization rotation occurs, the frequency is unchanged and the photonic qubit is always correlated with the correct atomic state. Under a larger birefringence $\delta = 2\kappa$, as shown in Fig.\,\ref{fig:2}(b), the birefringent component is comparable to the non-birefringent component in polarization encoding. For frequency encoding, the wavepackets maintain similar shape with little influence from birefringence.

We then simulate atom-atom entanglement fidelity after two-photon interference. 
Fig.\,\ref{fig:3}(a) shows the fidelity and efficiency as a function of coincidence window $T$ in the absence of birefringence. Here, the coincidence window $T$ represents the maximum allowed detection time difference for blue and red photons, $|t_r - t_b|$. For each $T$, the fidelity is averaged within the range $|t_r - t_b| \leq T$. Since re-excitation is ignored and both nodes are considered identical, the fidelity is always one, while the collection efficiency grows asymptotically, and is normalized such that it approaches one when $T$ is much larger than the wavepacket bandwidth ($\sim1/\kappa$). Fig.\,\ref{fig:3}(b) and (c) show fidelity as a function of the coincidence window. Two solid lines represent the result with the same birefringence $(\delta_A, \delta_B)=(\kappa,\kappa)$. In this case, the frequency encoding shows unit fidelity, whereas polarization encoding results in significantly lower fidelity. The dashed lines represent the result under an imbalanced birefringence $(\delta_A, \delta_B)=(0.5\kappa,\kappa)$. Here, the fidelity of frequency encoding drops because the photons from both nodes are no longer identical. On the other hand, the fidelity of polarization encoding improves compared to Fig.\,\ref{fig:3}(b), because the reduced birefringence at node A lowers the probability of producing a photon entangled with the incorrect atomic state.

Fig.\,\ref{fig:4} displays the windowed fidelity under varying birefringence magnitude at both nodes for both frequency encoding and polarization encoding. For the frequency encoding scheme, as shown in Fig.\,\ref{fig:4}(a), the fidelity may be unaffected even at large birefringence $\delta \sim \kappa$, provided that the birefringence in both nodes is the same. The fidelity only decreases when birefringence differs between nodes, reducing to 60\% when $\delta_A = 0$ and $\delta_B = 2\kappa$. This is because the birefringence in node B changes its wavepacket temporal shape relative to node A and breaks the photon indistinguishability. For the polarization encoding scheme, as shown in (b), high fidelity only appears in the region where birefringence at both nodes is sufficiently small. The fidelity is sensitive to the birefringence of each node individually, and decreases substantially with either of them increasing. Even under a small birefringence $(0.3\kappa,0.3\kappa)$, which corresponds to a minor birefringent component as shown in Fig.\,\ref{fig:2}(a), the fidelity drops to $\sim85$\%.

To summarize, birefringence impacts REG fidelity in two ways. It can directly reduce fidelity by causing photons to become entangled with incorrect atomic states, or different birefringence between nodes can affect photon indistinguishability, leading to further reductions in fidelity. While the polarization encoding scheme is sensitive to both aspects, the frequency encoding scheme is robust against the first one and is less affected when the amplitudes of birefringence are similar between the two nodes (Fig.\,\ref{fig:4}).

\section{Experimental Scheme}\label{sec:5}
In this section, we consider two experimental implementations of the frequency encoding scheme: using the atomic fine structure splitting of $\mathrm{^{40}Ca^+}$ and the hyperfine splitting of $\mathrm{^{225}Ra^+}$ ions. Both schemes can be adapted to other atomic species with a fine structure on metastable states or a hyperfine structure, such as $\mathrm{^{88}Sr^+},\mathrm{^{133}Cs},\mathrm{^{138}Ba^+},\mathrm{^{171}Yb^+},$ etc. Cavity and laser parameters used in our simulations are provided in Tab.\,\,\ref{tab:1}. We calculate the fidelities of  each realization of the frequency encoding scheme using realistic parameters in the presence of birefringence. 

\begin{figure}[t]
    \centering
    \includegraphics[width=0.48\textwidth]{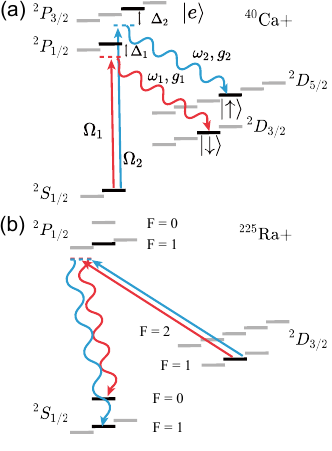}
    \caption{
   Relevant energy levels of $\mathrm{^{40}Ca^+}$and $\mathrm{^{225}Ra^+}$ for implementation of a frequency-encoded quantum network. (a) Implementation using the fine structure splitting of the metastable D state in $\mathrm{^{40}Ca^+}$. 
   (b) Implementation using the hyperfine splitting of the electronic ground state in $\mathrm{^{225}Ra^+}$. 
   }
    \label{fig:5}
\end{figure}

\begin{table*}[!t]
  \centering
\begin{tabular}{|c|c|c|c|c|c|c|c|c|c|c|c|c|c|}
  \hline
   Ion species&$l$/mm&$R_c$/mm&$\mathcal{F}$&FSR&$g_1$& $g_2$&$\kappa$&$\Omega_1$&$\Omega_2$&$\Delta_1,\Delta_2$&$\gamma_{ie}$& $\gamma_{xe}$ &$\delta$ \\
  \hline
    $\mathrm{^{40}Ca^{+}}$& 0.493 & 0.493 & $5\times10^4$ & $304\times 10^3$ & 9.36 & 8.89 & 6 & 38 & 40 & 400 & 10.78$\times$1/3 &7.92 & 0.5$\kappa$ \\
  \hline
   $\mathrm{^{225}Ra^{+}}$& 10.8 & 10.8 & $5\times10^4$ & $13.84\times 10^3$ & 1.01 & 1.01 & 0.28 & 40 & 40 & 400 & 1.69$\times$1/3 &9.00 & 0.5$\kappa$ \\
   \hline
\end{tabular}
   \caption{The parameters used for simulation. Values in columns $\mathcal{F}$, FSR, $g_1$, $g_2$, $\kappa$, $\Omega_1$, $\Omega_2$, $\Delta_1,\Delta_2$, $\gamma_{ie}$ and $\gamma_{xe}$ are given in the unit of  $2\pi\cdot\text{MHz}$. $\Omega_{1,2}$ and $\Delta_{1,2}$ represent the Rabi frequency and detuning for the two Raman drive lasers. $\gamma_{ie}$ is the spontaneous emission rate from $\ket{e}$ back to $\ket{i}$, and $\gamma_{xe}$ is the total spontaneous emission rate from $\ket{e}$ to all other levels. The atom-cavity coupling strength $g_1$ and  $g_2$ for the two CARTs and the cavity decay rate $\kappa$ are calculated according to cavity parameters: cavity length $l$, mirror radius of curvature $R_c$ and finesse $\mathcal{F}$. For $\mathrm{^{225}Ra+}$ ion, $l$ is designed such that the cavity FSR matches half the frequency difference between $\ket{S_{1/2},F=0,m_F=0}$ and $\ket{S_{1/2},F=1,m_F=0}$.}
    \label{tab:1}
\end{table*}

\begin{figure*}[!t]
    \centering
    \includegraphics[width=0.8\textwidth]{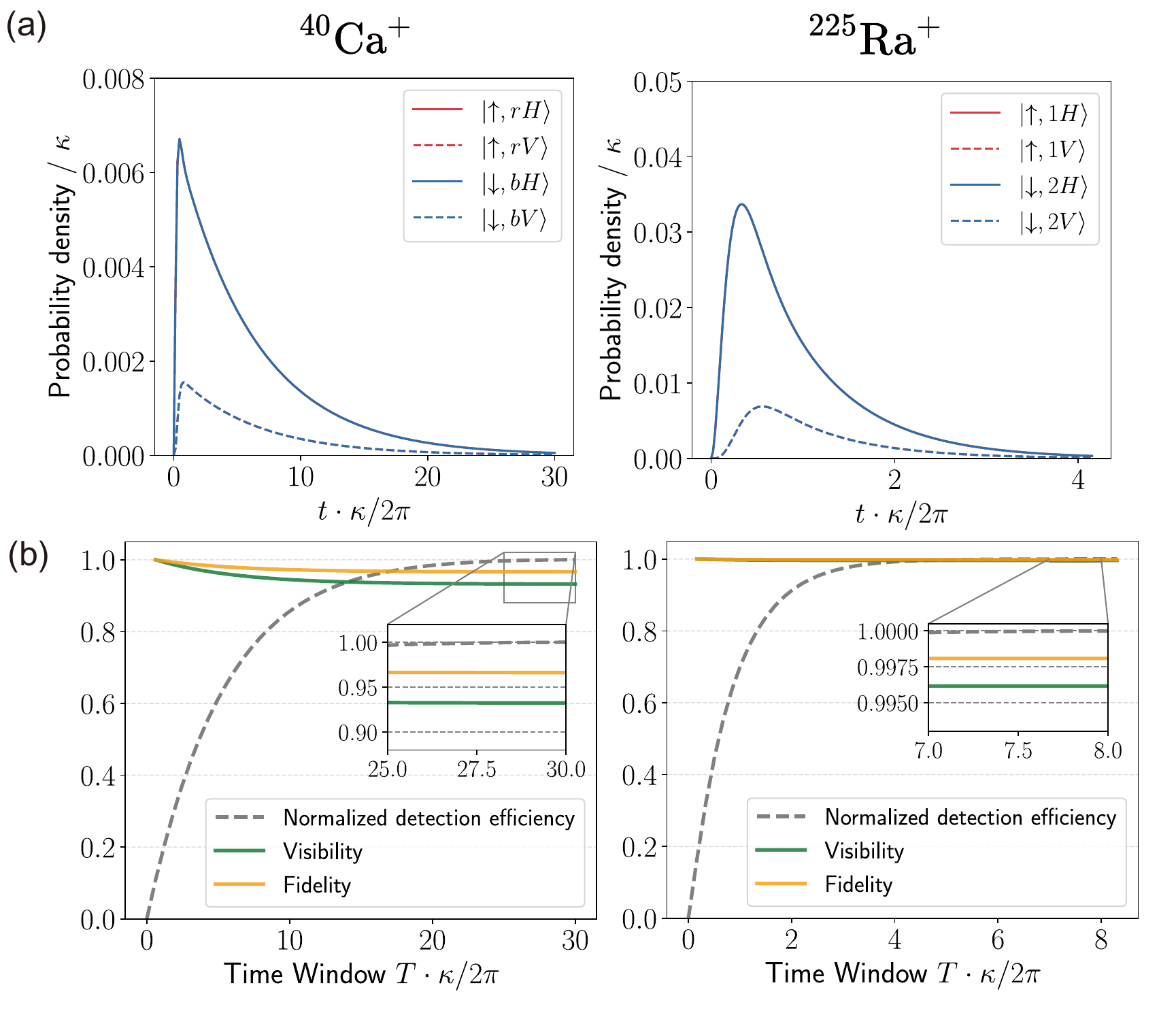}
    \caption{(a) Atom-photon state at the cavity output of the frequency encoding scheme using $\mathrm{^{40}Ca^+}$ and $\mathrm{^{225}Ra^+}$ scheme. The blue and red curves represent photons of different frequency states, which overlap and are not distinguishable in the figure. (b) Detection efficiency, visiblity of two-photon inference, and atom-atom fidelity
under of the two use cases. Identical parameter settings are used for both nodes, and re-exciation process is considered. Insets show the asymtoptic limit at long coincidence window.}
    \label{fig:6}
\end{figure*}

\subsection{Qubits with fine structure splitting}
We first consider an experimental realization using $\mathrm{^{40}Ca^+}$ because of its small fine-structure splitting. The atom is initially prepared at $\ket{i}=\ket{S_{1/2},m_J=+1/2}$. Two laser fields each couple non-resonantly to the  $\ket{P_{3/2},m_J=1/2}$ and $\ket{P_{1/2},m_J=1/2}$ states. Then the population is transferred to $\ket{\uparrow} = \ket{D_{5/2}, m_J = 1/2}$ and $\ket{\downarrow} = \ket{D_{3/2}, m_J = 1/2}$ respectively via a CART, emitting photons at wavelengths of $854$\,nm and $866$\,nm into the cavity. Transitions to other Zeeman sublevels are suppressed due to the frequency mismatch between the emitted photon and the cavity mode. After the atom-photon entanglement operation, $\ket{\downarrow}$ can be mapped to other Zeeman sublevels in the $D_{5/2}$ manifold using a Raman pulse to execute local quantum operations~\cite{lai2024realization}.

In our simulation, the parameters for the two nodes are set to be identical and the re-excitation process is characterized by the rate $\gamma_{ie}/(2\pi) = 10.78\times1/3$\,MHz~\cite{hettrich2015measurement}. With such a high decay rate, re-excitation errors will be a leading contributor to fidelity loss. To mitigate this effect while maintaining similar collection efficiency for pure photons, a large atom-cavity coupling strength $g$ can  limit population accumulation in the excited state. With these considerations, we model a fiber Fabry-P\'erot cavity with a Finesse of $\mathcal{F}= 5\times10^4$, cavity length $l=0.493$\,mm and mirror radius of curvature $R_c=0.493$\,mm. This gives a beam waist of  $w = (8.24,\,8.18)\,\mu\text{m}$, and  $g/(2\pi)=-(14.8\times\sqrt{10}/5,\,15.4\times\sqrt{3}/3)\,\text{MHz}$ for the 866\,nm and 854\,nm transitions, respectively. The second operands come from the Clebsch–Gordan coefficients. We also set the birefringence magnitude to be $\delta = 0.5\kappa$ and the cavity eigenbasis to be $(\ket{H}\pm\ket{V})/\sqrt{2}$. 

The simulation results are shown in Fig.\,\ref{fig:6}. Assuming a large cavity decay $\kappa/(2\pi)=6$ MHz, the wavepacket shape is dominated by the Rabi frequency of the driving light. For small coincidence windows, the windowed fidelity starts at 100\%, decreasing for a larger time window and eventually saturating at $96.6\%$ for longer ones, limited by re-excitation errors. 

\subsection{Qubits with hyperfine splitting}

Frequency-encoded photons can also be generated using atoms with hyperfine structure. We consider an experimental scheme utilizing the ground state of $\mathrm{^{225}Ra^+}$ with nuclear spin $I=1/2$ that has a large hyperfine splitting and a photonic qubit at 468\,nm. The atom is initially prepared at $\ket{i}=\ket{D_{3/2},F=1,m_F=0}$ and driven to $\ket{\uparrow}=\ket{S_{1/2},F=1,m_F=0}$ and $\ket{\downarrow}=\ket{S_{1/2},F=0,m_F=0}$ respectively via CART, mediated by excited state $\ket{e}=\ket{P_{1/2},F=1,m_F=-1}$. The two possible photonic states emitted in this process each have a wavelength near $468$\,nm with frequency difference $27.68\,\text{GHz}$~\cite{ready2024laser}.

One advantage of using $\mathrm{^{225}Ra^+}$ is its lower decay rate back to the initial state $\gamma_{ie}/(2\pi) = 1.69\times1/3$\,MHz, leading to much lower re-excitation infidelity than in the $\mathrm{^{40}Ca^+}$ scheme. However, due to the small frequency difference of just $27.68\text{\,GHz}$, the cavity FSR needs to be such that the cavity supports both frequency modes. To meet this demand, we model a confocal cavity with a length of $l=10.8$\,mm and mirror radius of curvature $R_c=10.8$\,mm. The beam waist in this case is $w = 28\,\mathrm{\mu m}$, and $g/(2\pi)= 1.75\times1/\sqrt{3}$\,MHz. We choose for both cavities the same birefringence ($\delta = 0.5\kappa$) and the cavity eigenmode ($(\ket{H}\pm\ket{V})/\sqrt{2}$) as above. The simulation results are shown in Fig.\,\ref{fig:6}. The windowed fidelity starts at $100\%$ and decreases to $99.8\%$ asymptotically for an arbitrarily long coincidence window.








\section{Conclusion}\label{sec:6}
We have presented a novel experimental scheme for a cavity-based quantum network using frequency-encoded photonic qubits. Compared to polarization encoding, this scheme is less sensitive to cavity birefringence, with its only contribution to infidelity originating from the reduction in indistinguishability resulting from differences in birefringence between the two nodes.

We performed a detailed numerical analysis comparing frequency and polarization encoding to validate the effectiveness of our proposed scheme. This analysis includes the temporal emission probability of photonic states, two-photon interference visibility, and an evaluation of the robustness in entanglement fidelity to cavity birefringence. Furthermore, we discuss two implementations that show that even with substantial common birefringence~\cite{takahashi2014novel} in the two nodes $\delta_1 = \delta_2 = 0.5\kappa$, fidelities of 96.6\% for $\mathrm{^{40}Ca^+}$ and  99.8\% for $\mathrm{^{225}Ra^+}$ can be achieved.

Frequency encoding in a cavity-based quantum network still requires further experimental validation, and we identify several implementation challenges. Firstly, a proper detection and manipulation protocol of frequency-bin photons~\cite{lu2020fully} should be carefully selected or developed to achieve ideal separation between bichromatic photons. Secondly, compensation for fiber dispersion may be necessary; otherwise, temporal distinguishability may be introduced if the difference in arrival times of the two colors of photon reaches the scale of their bandwidth. These effects rely heavily on the chosen qubit frequency, as a larger difference in frequency might simplify detection, but introduce more dispersion. Fiber-induced polarization drift should also be actively stabilized in all encoding schemes~\cite{knaut2024entanglement,uysal2024spin}, including frequency encoding. Additionally, frequency encoding does not guarantee high fidelity when the birefringence in both nodes is not identical. These errors can be further mitigated by optimizing the Raman laser and cavity parameters at the two nodes for maximal indistinguishability, including laser detuning, laser drive Rabi frequency, atom-cavity coupling strength, cavity decay rate, etc.

Our findings pave the way for a more robust design of cavity-based quantum networks, offering an alternate path towards achieving high-fidelity, scalable quantum entanglement over long distances. Together with experimental realization, our scheme can find wide application in quantum key distribution, quantum enhanced sensing, and distributed quantum computing.

\section{Acknowledgments}
We thank Wance Wang, Ezra Kassa and Yiheng Lin for their helpful discussion. Q.W. and H.H. acknowledge funding by the U.S. Department of Energy, Office of Science, Office of Basic Energy Sciences under Awards No. DE-SC0023277.
This work was supported by the Office of Advanced Scientific Computing Research (ASCR), Office of Science, U.S. Department of Energy, under Contract No. DE-AC02-05CH11231. The research was executed under the Quantum Internet to Accelerate Scientific Discovery ASCR Research Program funded through Berkeley Lab FWP FP00013429.


\bibliography{frequency_encoding}


\end{document}